# 3D Model Classification and Retrieval Based on Semantic and Ontology

My Abdellah Kassimi [1] and Omar El beqqali [2]

GRMS2I, Sidi Mohamed Ben Abdellah University,
Fez Morocco
[1]*abdellahkassimi@yahoo.fr*
[2]*omarelbeqqali@gmail.com*

**Abstract**
Classifying 3D models into classes is an important step in 3D model retrieval process. However, most classification system does not include semantic information. In this paper, a new method has been proposed to classify and to retrieve 3D model using semantic concepts and ontology. First, we use the machine learning methods to label 3D models by k-means algorithm in shape indexes space. Second, classification of 3D models into classes based on semantic annotation is applied. Finally, ontology is constructed with instances for each class, which contain spatial relationships, shape indexes and measures. Therefore, the 3D model retrieval system is comprised of low-level and high-level visual features. We interpret our results using the Princeton Shape Benchmark Database and our prototypical 3D Model Search System (3DMSS). The results show the performance of the proposed new approach to classify and to retrieve 3D models.

## 1. Introduction

The size of 3D models that is used on the web or stored in databases is becoming increasingly high. To increase the identification rate and decrease the time to search for items, different methods have been developed for the classification of 3D models, using geometrical characteristics. However, most of these methods do not include semantic information. Therefore, 3D model retrieval system has been affected by the similarity gap between the lower and the higher level features. In this paper, two ways are used to reduce the semantic gap: First, unsupervised learning method is used to create the link between shape indexes features and the semantic concepts. Then, the classification of 3D models which is based on semantic annotation is applied. Second, using OWL ontology to define the concepts of 3D models, the spatial relationships are applied to disambiguate among models of similar appearance. 3D model retrieval system which is based on the semantic and ontology is developed through the use of 3D shape indexes and spatial relationships represented by concepts in ontology. There are two motivations for using shape indexes: For the first motivation, shape indexes, which are all normalized, are frequently used to quantify different aspects of 3D model shape. Concerning the second motivation is to extract visual concepts easily, and semantic information can be extracted using unsupervised learning method.

## 2. Related work

Several systems and approaches for the classification of 3D models have been proposed in the literature. Chin-Chia Wu and al. [1] proposed the new approach for classifying 3D models in points clouds based on geometric graph representation. The approach uses a RIMLS technique and spin image signature to calculate the geometric characteristic. Based on the spatial clustering ontology, the authors in [3] developed ontology-based spatial clustering and reasoning system. This system integrates domain knowledge and user goals into clustering. Maria and Mihai [2] proposed the classification method based on clustering algorithm for body shape recognition. For the 3D model-semantic problem, many approaches have been proposed. The work presented in [4] introduces the classification and retrieval 3D model by integrating shape features and semantic information. The paper proposes a new type of shape feature based on 2D views and use Gaussian processes as supervised learning to mode the mapping from low-level features to query concepts. In the paper [5], the author explores a new framework for 3D model retrieval based on an off-line learning of the most salient features of the shapes. The proposed approach uses a large set of features, which can be heterogeneous, to capture the high-level semantic concepts of different shape classes. Hou and al, in [6] Support Vector Machine (SVM) is used to cluster 3D models with respect to semantic information for the organization a database of shapes. Therefore, the classification and retrieval 3D model system is integrated [7]. Also the semantic description of an object based on the ontology, the matching of this description with the low-level features such as color, texture, shape and spatial relationships [8] [9] are used to classify and indexing images. In paper [10], authors incorporate semantics provided by multiple class labels to reduce the size of feature vector produced by bag-of-features [11] exploiting semantics.

In this paper, we suggest reducing the semantic gap in two steps during 3D model retrieval process: the first one aims at classifying 3D models based on semantic concepts. To label 3D models, in this step, the machine learning methods is applied in shape indexes space. Second, we use SPARQL engine to question the information displayed in OWL ontology using spatial relationships.

## 3. System overview

In this paper, our content-based search (3DMSS) is used to test the proposed classification and retrieval. The proposed 3DMSS is illustrated in figure ("Fig 1").

In the inline process, the user can navigate in the database and sends a 3D request to the server. The 3D model retrieval procedure is a three-step process: first, 3DMSS defines the class membership which is semantically classified. Second, sends the request to ontology by SPARQL engine using spatial relationships. Finally, compares its descriptor with the descriptors of all models selected in second step using geometrical characteristics.

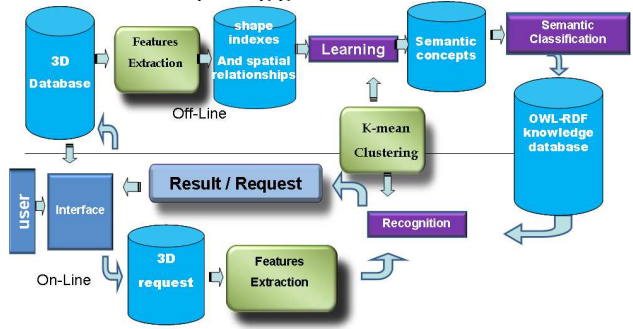

Fig 1. Overview of the proposed system

Our 3D Database is composed of Princeton Shape Benchmark 3D models [12] that are stored in a format (*.off) which represents the shape of 3D models by polygonal mesh.

## 4. Descriptor Extraction

The size and shape of 3D model have been used to describe 3D model. There are several ways to describe a 3D model shape. However, no single shape descriptor is appropriate for all models. Therefore, such a way of characterizing shape is required. Shape indexes that provide Sphericity, Compactness, Convexity and elongation, are frequently used to quantify different aspects of 3D model shape ("Fig 2"). In addition, motivation for using shape indexes is extracting visual concepts easily and tells a lot about the semantics of the 3D model. Therefore, the association of shape indexes with semantic concepts should be used.

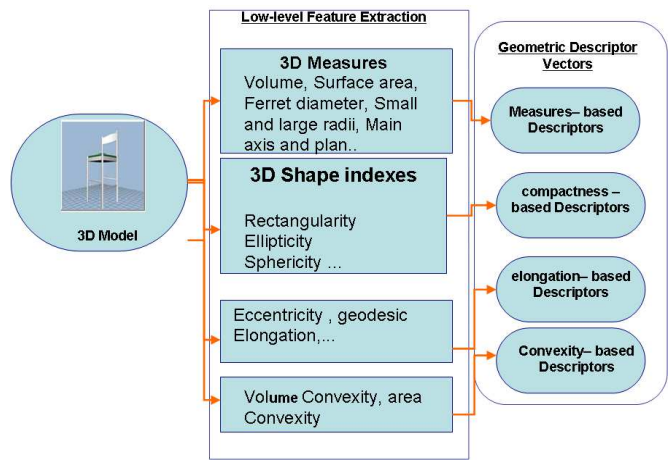

Fig 2. Descriptor Extraction

To describe 3D model size, various measures from this 3D model are calculated. The most frequently used are the diameter and the length measure of the three representative main axes. The measure equivalent spherical diameter (ESD) can be used to define 3D model size. It is the diameter of a sphere of equivalent volume that gets larger or smaller as the model does. Let $S_{ch}$ the convex hull surface area. The ESD is defined by:

$$ESD = (4/3) * \pi * \sqrt{(S_{ch}/\pi)^3} \qquad (1)$$

To compute 3D shape indexes, the most important 3D measures are: surface area and volume. With 3D polygonal model representation, these measures [15] can be readily obtained from the models and are used directly for calculating 3D shape indexes. For other 3D measures, we use the 2D measures transformed by modifying the calculation. In practice, we used the following measures: Volume, Surface area, diameter, Ferret diameter, Small and large radii, main axis, plan and ESD. Similar measures are calculated from bounding box and convex hall that is the minimum enveloping boundary. These measures are used as semantic concepts in ontology and allow to define the spatial relationships.

From these basic measures, multiple shape indexes are calculated for each 3D model and subdivided into four groups: Compactness, Sphericity, Convexity and Elongation. For others shape indexes see our works in [14].

### 4.1 Compactness

From ESD measure, new shape index that provides compactness indicator can be calculated [9] as follows:

$$C1 = ESD / D_{ch} \qquad (2)$$

Where $D_{ch}$ is the convex hall diameter.

Based on volume (V) and surface area (S) ratio, other shape index characterizing the Compactness of 3D model have been calculated as follows:

$$C2 = (V^2 * 36 * \pi) / S^3 \qquad (3)$$

### 4.2 Sphericity

Sphericity and roundness have been used to represent the 3D model shape and are indications of compactness. Sphericity is a shape index that provides how spherical an object is. Roundness is related to angularity and represents the curvature of model's corners. New shape index characterizing the Sphericity of 3D model has been calculated [24] as follows:

$$S1 = a * V^{2/3} / S \qquad (4)$$

Where $a = 6^{2/3} \pi^{1/3} \cong 4.84$ (which makes the S1 equal to 1 for a sphere).

The Sphericity index is a dimensionless constant with values ranging from zero for a laminar disc to unity for a sphere and it is most sensitive to elongation.

### 4.3 Elongation

Elongation provides an indication of 3D model overall form by comparing the strength of the major axis and the strength of the minor axis of a 3D model.

It is defined as (1-[width/length]) and has values in the range of zero to one. We are using main axes and radii to compute elongation. We can use also moments to compute them as in [16].

### 4.4 Convexity

This shape indexes group computes and provides the surface roughness of a 3D model and is calculated by dividing the convex hull surface area by a surface area of 3D model [14].

$$C_s = S_{ch} / S \qquad (5)$$

We can also calculate convexity shape index using 3D model volume as follows:

$$C_v = V_{ch} / V \qquad (6)$$

A smooth shape, regardless of form, has a convexity of 1 while a very 'spiky' or irregular object has a convexity closer to 0.

Shape indexes calculated are quick to compute, easy to understand and were chosen mostly for their simplicity and are invariant to rigid motions such as translations and rotations. However, the important idea is extracting semantic concepts easily from shape indexes to classify 3D models. Measures and shape indexes are considered descriptors in this paper and others definitions are detailed in [13] [14].

## 5. Semantic-based classification

The next step after features extracting is to classify 3D models semantically. The problem is assigning an appropriate class to the query model. The proposed 3D model classification contains a semantic labeling step and a classifying step. In first one, we exploit shape indexes for semantic labeling and we use machine learning methods to associate shape indexes with semantic concepts as shown in figure 3. In this paper, shape indexes are used to represent 3D model visual concepts [17].

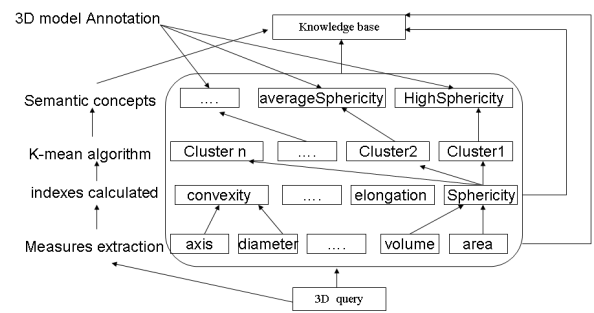

Fig 3. Definition of semantic concepts and Knowledge base augmented and guided by a 3D Shape index ontology to describe the 3D Models

Measures and shape indexes are clustered by a k-means algorithm into semantic clusters. The notion of similarity is based on each category of 3D shape indexes or measures like in figure ("Fig 3"). This approach is divided into the following steps: measure extraction; clustering and definition of semantic concepts. From the 3D database, the three steps are repeated for each 3D shape index to define semantic concepts. Therefore, 3D model is described by a set of the numerical value associated with semantic concepts.

The second step is classifying 3D models based on these semantic concepts. In general, classification is done after training. In this paper, k-nearest neighbor algorithm (k-NN) is used as a method for classifying 3D models. Based on closest training examples in the feature space ("Table 1"), k-NN classifies an object by a majority vote of its neighbors, with the object being assigned to the class most common amongst its k nearest neighbors. Euclidean distance is used as the distance metric. For example, Sphericity and Ellipticity IDs are based on the following table:

Table 1.  Semantic concepts IDs

| ID | Ellipticity | Sphericity |
|---|---|---|
| 0 | "High Ellipticity" | "High Sphericity" |
| 1 | "Average Ellipticity" | "Average Sphericity " |
| 2 | "small Ellipticity" | "small Sphericity " |
| 3 | "smaller Ellipticity" | "smaller Sphericity " |

In this example, the semantic concept is assigned to ID in semantic labeling step and applied to all semantic concepts. We should create a database to describe all models by the semantic concepts guided by a 3D Shape indexes ontology and relations among entities. The ontology defines a database structure as containing of a set of concepts that can describe qualitatively the visual semantic concepts and should allow similarity searches.

## 6. Ontology-based retrieval

### 6.1 Ontology

After building the classes based on semantic, the next step to reduce the semantic gap is constructing the ontology for each class. Ontology is employed to organize semantic concepts (e.g. Sphericity, elongation, convexity...) and other concepts such as semantic entities (e.g. lines, points, surface, and plan). This Ontology also comprises set of spatial relations and some axioms (transitivity, reflexivity, symmetry). The proposed ontology is represented in Ontology Language OWL [18].

SPARQL is used to request this ontology and the result is considered as the second method for the classification and the selection of 3D model in one class membership. Finally, to evaluate the similarity between two 3D models in all models selected based on visual concepts; appropriate distance by numeric value is required. In this paper, measures, shape indexes and moments (section 6.2) are used as numeric value to compute similarity between two 3D models using Euclidean distance.

### 6.2 Spatial relationships

To describe spatial relationships that are usually defined according to the location of the measure in the 3D model, we calculated the local characteristics from convex hull and bounding box. Therefore, four points from convex hull are considered ("Figure 4"): centroid (P1), the closest point to centroid (P2), the farthest point to centroid (P3), and the farthest point to P3 (P4). In the first step, the center of gravity G (or P1) is located. Then, we calculate other points using Euclidean distance.

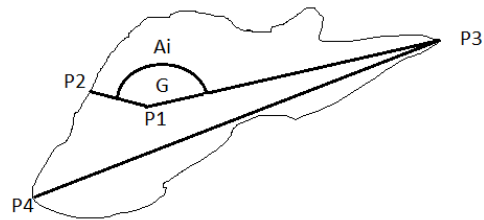

Fig 4.  Four points to describe spatial relationships

The maximum P1P3 and minimum distance P1P2 (respectively Small and large radii) are used to define the bounding box and there ratio is used as a dimensionless shape index.

From these points, we can compute polygons [23], lines, plans. Then, ratios computed between surface area and volumes from polygons are inserted in ontology as relationships. For each point Pi we calculate also the angle Ai, this angle allow to compute various directions. This set of features allows the description of model independently of their size, rotation, translation or line type [23]. During the process of calculating the four sets of point distances (P1, P2, P3, P4), the moments of distributions are calculated (12 moments of the resulting four distributions) [22] and are used to compute similarity via numeric value.

From bounding box three main axes are considered to describe position, distances and orientation of an entity in the 3D model. Therefore, several relationships are described ("Figure 5"):

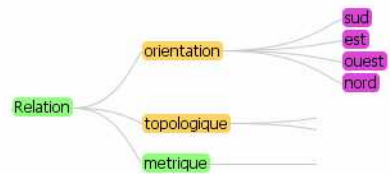

Fig 5.  Partial hierarchy of relationships

The distances can be computed from a point to point, line to line, point to line, point to plan and line to plan. Also we are interested in topological relationships among entities that are related to how objects interconnect. In this paper, we adopt the topological relationships as shown in table ("Table 2"). The RCC-8 [19] [20] relations can be used for taking into account spatial relations. RCC (Region Connection Calculus) is a logic-based formalism symbolically to represent and reason with topological properties of objects [21].

Table 2. Topological relations implemented in our system

| Point-Point | Point-Line | Line-Line | Line-Plan |
|---|---|---|---|
| ● | ●— | ✕ | □● |
| Overlap | On | Cross | Contained |
| ●● | ●— | ‖ | □● |
| Adjacent | Adjacent | Not Cross | Adjacent |

Based on the spatial relationships and their properties, we build the ontology using the web ontology language (OWL).

## 7. 3D model retrieval

The third step in our 3D model retrieval system is to compute de similarity between two 3D models as shown in figure ("Figure 6").

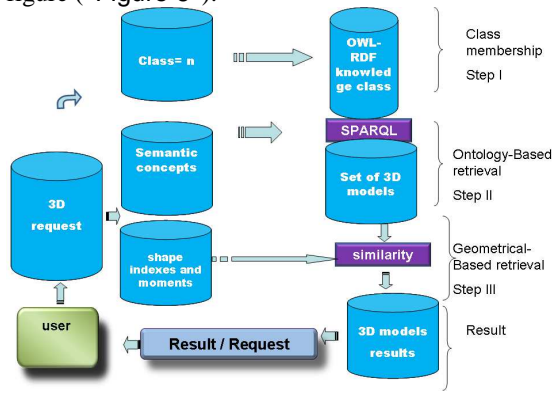

Fig 6. semantic and numeric query to evaluate the similarity

The similarity between two models is measured through the use of a distance between their measures, 3D shape indexes and the moments of distributions calculated during the process of calculating the four sets of point distances (P1, P2, P3, and P4).
In our system, the Euclidean distance is used to measure the similarity between 3D models. Therefore, to provide the best results, it is necessary to combine shape indexes, measures and moments to compute the most relevant. A simple approach for combining these descriptors is to compute the weighted sum of the distances [14].

## 8. Experimental Results

We are using Java language to develop our content-based retrieval systems for 3D models [14]. The average times that are used to compute all measures, shape indexes, moments and relationships is 0,114 seconds for a model, using the Princeton Shape Benchmark Database ("Table 3").

Table 3. Time to compute all descriptors

| Example model | Number of edges | number of polygons | Number of vertices | Time |
|---|---|---|---|---|
| 1 | 1638 | 546 | 341 | 0,128405697 |
| 2 | 648 | 228 | 100 | 0,131520481 |
| 3 | 1476 | 492 | 298 | 0,085323701 |
| 4 | 1224 | 408 | 216 | 0,234916242 |
| 5 | 40626 | 13542 | 7056 | 0,80845978 |
| 6 | 3336 | 1112 | 557 | 0,222504762 |
| 7 | 50925 | 16975 | 8469 | 0,774157548 |
| 8 | 5637 | 1879 | 781 | 0,24777793 |

During the process on line, all features are computed, and we can directly retrieval models as shown in figure 7 and 8 ("Fig 7, Fig8").
The 12 most similar models are extracting and return to do a user by 2D images. To visualize the 3D models in the 3D space, the user clicks the button or image.

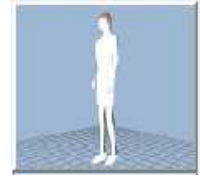

Fig 7. query model

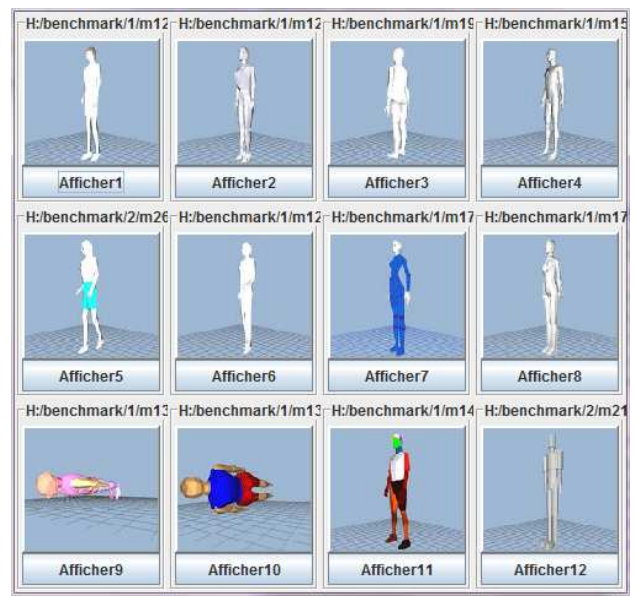

Fig 8. Models found with our visual descriptor

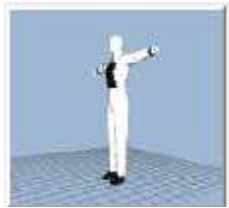

Fig 9.     query model

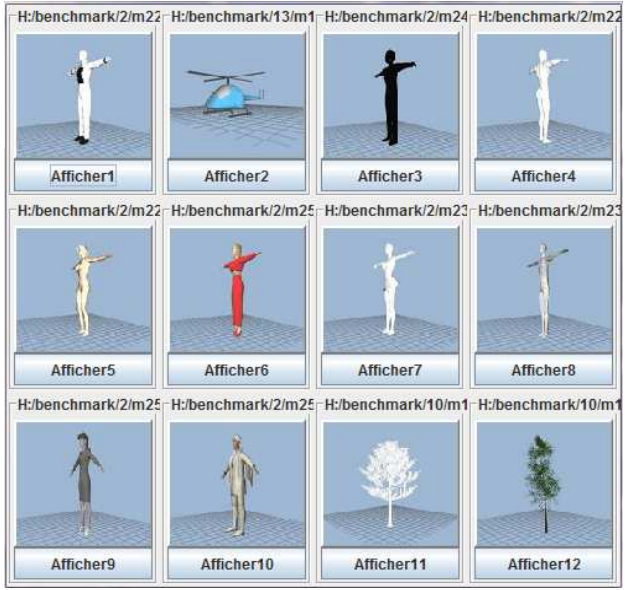

Fig 10.    Models found without classification without ontology

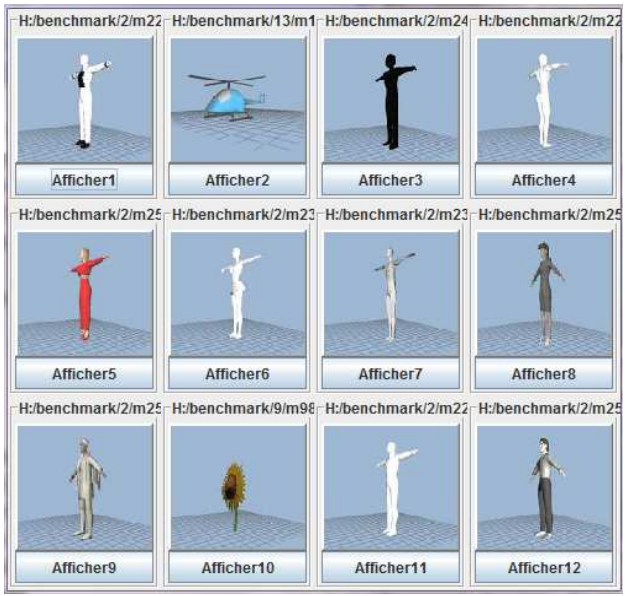

Fig 11.    Models found with classification and without ontology

To classify 3D models by introducing the semantic descriptor ("Fig 10, Fig 11, Fig 12"), the query is labeled before the search happens with a semantic concept by associating 3D shape low-level features with high-level semantic of the models.

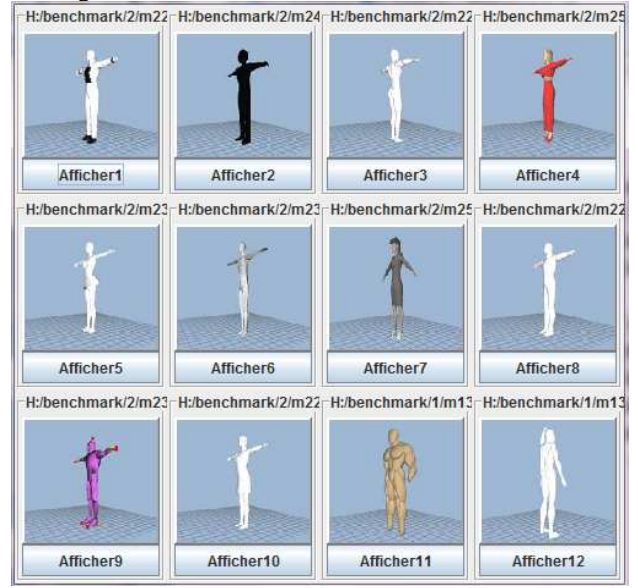

Fig 12.    Models found with classification and ontology

Figure ("Fig 11") and other examples show that our 3D classification based on semantic concepts is the important method to classify 3D model justifying the importance of using the shape indexes and measures as visual concepts.

The evaluation of our system consists of two main steps: shape indexes effectiveness and 3D model retrieval. For the first one, we evaluated the effectiveness of integrating new shape indexes to label 3D model and we compared the retrieval performances of the shape indexes at different groups. In this experiment, combining Sphericity, convexity and elongation gives the most reliable results.

Concerning second step, 3D Harmonics and Moments are implemented. We used the Recall and Precision to compare different descriptors. Figure ("Fig 13"). shows that our proposed descriptor performs well.

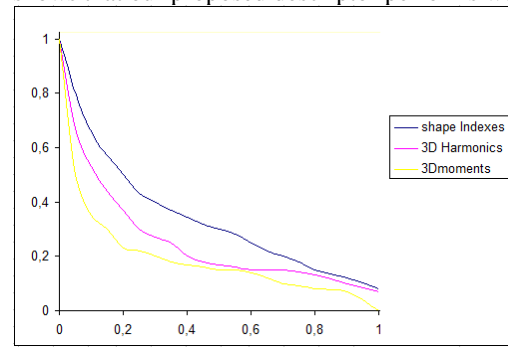

Fig 13.    The precision-recall curves of our system

The classification based on semantic concepts developed reduces the similarity gap, and the retrieval method by introducing the ontology retrieval is considered as the second method to classify 3D model in one class membership. Both methods to classify 3D models allow our system more efficient. This performance is also linked to the combination of shape indexes and semantic concepts structured in ontology.

## 9. Conclusion

A new method for 3D models classification and retrieval is introduced in this paper. First, the 3D classification that has been based on semantic concepts is proposed. Second, the method combines semantic concepts and geometrical characteristics which are structured in ontology to 3D model retrieval. The new approach is tested with a large 3D database using the search engine developed, which allows us to show the relevance of our method. The results are promising and show the interest of our approach.

**My abdellah Kassimi** is a PhD student at Sidi Med Ben AbdEllah University (GRMS2I group) in Morocco. He received his DESS in Computer Science from the University of Sidi Md Ben AbdEllah in 2007. His current research interests are 3D indexing and retrieval, 3D shape indexes, semantic and ontology.

**Omar El Beqqali** is currently Professor at Sidi Med Ben AbdEllah University. He is holding a Master in Computer Sciences and a PhD respectively from INSA-Lyon and Claude Bernard University in France. He is leading the 'GRMS2I' research group since 2005 (Information Systems engineering and modeling) of USMBA and the Research-Training PhD Unit 'SM3I'. His main interests include Supply Chain field, distributed databases and Pervasive information Systems. He also participated to MED-IST project meetings. O. El Beqqali was visiting professor at UCB-Lyon1 University, INSA-Lyon, Lyon2 University and UIC (University of Illinois of Chicago). He is also an editorial board member of the International Journal of Product Lifecycle Management (IJPLM).